\documentclass[12pt]{article}
\usepackage[latin1]{inputenc}
\usepackage[T1]{fontenc}
\usepackage{amsmath}
\usepackage{amsfonts}
\usepackage{amssymb}
\usepackage{color}
\usepackage{graphicx}
\usepackage{placeins}
\usepackage{tikz}
\usepackage{graphicx}
\usepackage{caption}
\usepackage{subcaption}
\usepackage{xspace}
\usepackage[normalem]{ulem}
\usepackage{mathtools}

\newcommand{\be}{\begin{equation}}
\newcommand{\ee}{\end{equation}}
\newcommand{\ba}{\begin{eqnarray}}
\newcommand{\ea}{\end{eqnarray}}

\newcommand{\bea}{\begin{eqnarray}}
\newcommand{\eea}{\end{eqnarray}}

\newcommand{\beq}{\begin{equation}}
\newcommand{\eeq}{\end{equation}}

\newcommand{\fs}
{i\kern+.01em\hbox{\raise.20ex\hbox{$/$}\kern-.58em$s$}}
\newcommand{\dslash}{\!\!\not\!\partial}

\newcommand{\bs} {i\kern-.01em\hbox{\raise.25ex\hbox{$/$}\kern-.52em$b$}}
\newcommand{\qs}{/\kern-.52em s}

\newcommand{\dd}
{\kern.06em\hbox{\raise.25ex\hbox{$/$}\kern-.60em$\partial$}}

\begin{document}

\tikzstyle{bag} = [text width=2em, text centered]
\tikzstyle{bag1} = [text width=5em, text centered]
\tikzstyle{end} = []
\title{
A $d=3$ dimensional model with two $U(1)$ gauge fields coupled via matter fields and BF interaction
}
 \author{A.~Rapoport$^a$   and
F.~A.~Schaposnik$^b$
\\
~
\\
~
\\
{}\\
{\normalsize \it $^a$\!Departamento de F\'{\i}sica, FCEYN Universidad de Buenos Aires \& IFIBA CONICET,}\\
{\normalsize \it  Pabell\'on 1 Ciudad Universitaria, 1428 Buenos Aires, Argentina}
{\normalsize $\!$\it }\\
~
\\
{\normalsize $^b\!$\it Departamento de F\'\i sica, Universidad
Nacional de La Plata} \&  {\normalsize\it IFLP-CONICET,}\\
{\normalsize\it C.C. 67, 1900 La Plata,
Argentina}
 }

\date{\normalsize\today}

\maketitle
\begin{abstract}
We study a model in \mbox{$d=2+1$} space-time dimensions with two sectors. One of them, which can be considered as the visible sector, contains just a $U(1)$ gauge field which acts as a probe for the other (hidden) sector, given by a second $U(1)$ gauge field and  massive scalar and Dirac fermions. Covariant derivatives of these matter fields and a BF gauge mixing term couple the two sectors. Integration over fermion fields leads to an effective theory with Chern-Simons terms that support vortex like solutions in both sectors even if originally there was no symmetry breaking Higgs scalar in the visible sector. We study numerically the solutions which correspond to electrically charged magnetic vortices   except for a critical value of the BF coupling constant at which solely purely magnetic vortices exist.
\end{abstract}

\section*{Introduction}

Theories in which there are two gauge fields coupled via a gauge-mixing term, originally proposed in ref.~\cite{Holdom}, have recently received much attention both in high energy physics and in condensed matter. In the former case the idea is to construct models that account for the possible existence of dark matter composite sectors and also to implement supersymmetry breaking (see \cite{uno}-\cite{JR}-\cite{AriasNf}). Concerning condensed matter, this kind of models have been used to describe the so-called intertwined orders in high-temperature superconductors \cite{Ed}-\cite{Co} and also to study topological insulators and superconductors \cite{Sen}-\cite{SSWW}.

Fields in the dark matter hidden sector are in general taken as (complex) scalar fields and gauge fields (see for example \cite{f3} and references therein) but there are also very interesting proposals in which fermion fields play a central role as WIMPs, as for example to calculate the effective number of neutrinos species \cite{Wein0}. Of course, in supersymmetry breaking models the inclusion of fermions is mandatory (see \cite{Arias} and references therein). Other relevant cosmological effects in which fermionic fields in the hidden sector play a central role have been studied in \cite{Gru}.

In general, the coupling constants that mix fields of the two sectors should be taken much smaller than those in the observable sectors. This is the case, for example, concerning the possible existence of millicharged particles with an electric charge $\epsilon e$ for small $\epsilon$ \cite{JR}, being $e$ the elementary charge in the visible sector. In fact, recent experimental studies of the previously unexplored range $10^{-3} \leq \epsilon \leq 10^{-1}$ have increased the exclusion region for millicharged particles with masses greater than $0.1$ GeV and $\epsilon < 10^{-1}$ ref.~\cite{Argo}.

Remarkably, related ideas to those described above were put forward in the field of condensed matter physics to describe the complexity of phase diagrams in certain highly correlated electronic materials. Indeed, within a phenomenological Ginzburg-Landau superconductivity approach, distinct broken-symmetry phases associated to different competing orders were studied in models with several complex scalars \cite{Ed}-\cite{Frad}. There have also been interesting results concerning multicomponent superconducting systems. In particular it has been shown in ref.~\cite{Ba} that in a generic two component superconductor model (i.e. with two component scalars) a novel type of thermodynamically stable vortices with very unconventional magnetic properties can be found, which are relevant to the description of two-band superconductors. Progress in the analysis of dualities in \mbox{$d=3$} has also been applied in connection with materials in which topological order plays a central role (see for example \cite{Sen}).

In the present note we propose a model in \mbox{$d=2+1$} space-time dimensions in which the ``hidden'' sector includes a $U(1)$ gauge field and matter fields (fermions and a scalar) that are coupled to a $U(1)$ external gauge field belonging to the ``visible'' sector. As it is well-know, at high temperatures a 4-dimensional quantum field theory becomes effectively 3 dimensional. In particular, certain effective $d=4$ gauge theories coupled to fermions lead, in the high temperature limit \cite{Wj}-\cite{JaLibro}, to a residual Chern-Simons term. Hence, the $d=3$ effective theory that we discuss in this note could be of interest in connection with theories in $d=4$ at high temperature in which an external gauge field can be taken as a probe for the ``hidden'' sector. The coupling of the hidden sector with the external visible field is provided by covariant derivatives of matter fields with a hidden charge and also by a gauge mixing BF portal.


\section*{The model}

As stated in the introduction, we shall consider a model in \mbox{$2 +1$} space-time dimensions with metric ${\rm diag}\,g_{\mu\nu}=(1,-1,-1)$, having a ``visible'' and a ``hidden'' sector. Concerning the former, it consists of the observable $U(1)$ gauge field $A_\mu$, which can be considered as an external field. As for the hidden sector, it is composed of a complex scalar field $\phi$ with a symmetry breaking potential, a $U(1)$ gauge field $C_\mu$ and two massive Dirac fermions $\psi^i$ ($i=1,2$) coupled to $A_\mu$ and $C_\mu$ respectively.

The dynamics of the model is governed by the action ${\cal S} $
\bea
{\cal S} &=& \int d^3x \biggl(\frac{1}{2} \vert D_\mu[A]\phi\vert^2 - V[\phi] + \overline{\psi^1}\not\!\!D[A]\psi^1 -m_1\overline{\psi^1}\psi^1 + \overline{\psi^2}\not\!\!D[C]\psi^2 - m_2\overline{\psi^2}\psi^2 \\
&& + \frac{\kappa}{4\pi} \varepsilon^{\mu\nu\alpha} A_\mu \partial_\nu C_\alpha \biggr) \ .
\label{10bis}
\eea
Note that the only observable field coupled to the hidden sector is $A_\mu$.

The presence in the dark sector of a $U(1)$ gauge boson  and  a scalar  breaking this symmetry is a key ingredient in the proposal of a dark matter theory as discussed in ref.~\cite{Hamed}. In our proposal, the complex scalar $\phi$ couples to the visible gauge field $A_\mu$,
\begin{equation}
D_{\mu}[A]\phi = (\partial_{\mu} + i\epsilon eA_{\mu})\phi \quad , \quad \phi = \phi^1 + i\phi^2 \ ,
\label{2}
\end{equation}
and this will have relevant consequences as we shall see below.

Concerning the remaining terms of the first line in Eq.~\eqref{10bis}, the hidden fermions $\psi^i$ ($i=1,2$) are each one minimally coupled to the visible and the hidden gauge fields $A_\mu, C_\mu$,
\be
	\not\!\!D[A]\psi^1 = (\dslash + i \epsilon e\not\!\! A ) \psi^1 \quad , \quad \not\!\!D[C]\psi^2 = (\dslash + i e_h\not\!\! C) \psi^2 \ .
\label{3}
\ee
Here, $e$ represents the elementary charge in the visible sector. The coupling between hidden matter fields and the visible gauge field corresponds to a charge $\epsilon e$. If we take $\epsilon \ll 1$, then $\epsilon e$ can be interpreted as a milli-charge \cite{R}, as discussed in the introduction. As for $e_h$, it is the charge of the hidden fermion $\psi^2$.

The symmetry breaking potential $V[\phi]$ takes the form
\be
	V[\phi] =\frac{\lambda}{4}\left(\vert\phi\vert^2 - {\phi_0}^2\right)^2 \label{pot} \ .
\ee
Finally, the last term in the action \eqref{10bis} corresponds to a BF term coupling the visible and the hidden gauge fields.

As for the fields and parameters mass dimensions, one has
\[
	[\phi] = m^{1/2} \quad , \quad [A] = [C] = [\psi^i] = m \quad , \quad [e] = [e_h] = [\epsilon] = [\kappa] = m^0 \quad , \quad [\lambda] = m \ .
\]
Electric and magnetic fields are
\be
	B_A = -F^{12}[A] \quad , \quad E^i_A = -F^{i0}[A] \ ,
\ee
and analogous formul\ae\xspace hold for $C_\mu$ field.

\subsection*{The effective action}

As it is well known, each fermion determinant in $d=3$ dimensions has a parity odd and a parity even contribution. The parity odd one is the well-honoured Chern-Simons term \cite{Redlich}
\begin{multline}
\left. \int \mathcal{D}\bar\psi^{\,1} \mathcal{D}\psi^1 \exp\left(-\int d^3x {\bar\psi}^{1} \left(\not\!\! D[A] + m_1\right) \psi^1 \right)\right|_{\text{odd}} = \\ \exp\left(i\frac{(\epsilon e)^2}{8\pi}\int d^3x \left(\pm 1 + \frac{|m_1|}{m_1} \right) \varepsilon^{\mu\nu\alpha} A_\mu\partial_\nu A_\alpha \right)
\label{doble}
\end{multline}
and an analogous formula holds for the $\psi^2$ contribution, namely
\begin{multline}
\left. \int \mathcal{D}\bar\psi^{\,2} \mathcal{D}\psi^2 \exp\left(-\int d^3x {\bar\psi}^{\,2} \left(\not\!\! D[A] + m_2\right) \psi^2 \right)\right|_{\text{odd}} = \\ \exp\left(i\frac{e_H^2}{8\pi}\int d^3x \left(\pm 1 + \frac{|m_2|}{m_2} \right) \varepsilon^{\mu\nu\alpha} A_\mu\partial_\nu A_\alpha \right)
\label{doble2}
\end{multline}
with $e_h$ and $m_2$ hidden charge and mass respectively. Note that in Eqs.~\eqref{doble}-\eqref{doble2}  we have used the gauge invariant results obtained in \cite{GRS} using the $\zeta$-function regularization. Within this method, the origin of the double sign has to do with the   possible extensions of the complex powers of the Dirac operators. Indeed, depending on the choice of upper or lower half plane, the results differ in a  factor $(-1)^{d}$, with $d$ the space-time dimensions. Hence the ambiguity is always present in any odd dimensional case. Within Pauli-Villars regularization   the ambiguity is related to the fact that mass terms, the physical  and the regulator ones, violate $\mathcal P$ and $\mathcal T$ invariance. Finally, note that  Eqs.~\eqref{doble}-\eqref{doble2} can be also used in the $\lim_{m\to 0}$ case, still giving a non-trivial gauge-invariant result. A thoroughly discussion on these issues can be found in ref.\,\cite{GRS}.

As for the even parity contribution to the fermionic determinant, after an expansion in $\partial/m_1$ we obtain, to the leading order,
\be
\left. \int \mathcal{D}\bar\psi^{\,1} \mathcal{D}\psi^1 \exp\left(-\int d^3x {\bar\psi}^{1}\left(\not\!\! D[A] + m_1\right)\psi^1 \right)\right|_{\text{even}} = \exp\left(i\frac{(\epsilon e)^2}{48\pi \vert m_1\vert}\int d^3x F_{\mu\nu}[A]F^{\mu\nu}[A]\right) \ ,
\label{gaoF}
\ee
and an analogous result holds for $\psi^2$. We then have for the effective Lagrangian associated to action \eqref{10bis} after the fermionic path-integration,
\begin{equation}
\begin{split}
L_{\text{eff}} = &-\frac{\epsilon^2 e^2}{48\pi \lvert m_1\rvert}F_{\mu\nu}[A]F^{\mu\nu}[A] + \frac{1}{2} \vert D_\mu[A]\phi\vert^2 - V[\phi] \\
&+ \frac{1}{4\pi} \varepsilon^{\mu\nu\alpha} \left(\epsilon^2 e^2 A_\mu\partial_\nu A_\alpha +
 e_h^2 C_\mu\partial_\nu C_\alpha + \kappa C_\mu\partial_\nu A_\alpha\right) \ ,
\label{10vis}
\end{split}
\end{equation}
where we dropped the kinetic term for $C_\mu$ and assume for simplicity that $\lvert m_2\rvert \gg \lvert m_1\rvert$. From here on we shall consider positive fermion mass $m_1$ and comment at the end on the negative mass case. As for the sign indetermination, we chose to work with the positive sign (the result \eqref{doble} was obtained using a $\zeta$-function regularization in which the double sign arises in odd-dimensional theories).

We now introduce a field $\tilde{C_\mu}$ defined as
\be
\tilde{C}_\mu = C_\mu + \frac{\kappa}{2 e_h^2} A_\mu \ ,
\label{resta}
\ee
in terms of which Lagrangian \eqref{10vis} reads
\begin{equation}
\begin{split}
L_{\text{eff}} =& -\frac{\epsilon^2 e^2}{48\pi m_1} F_{\mu\nu}[A]F^{\mu\nu}[A] + \frac{1}{2} \vert D_\mu[A]\phi\vert^2 - V[\phi] \\
&+ \frac{1}{4\pi} \varepsilon^{\mu\nu\alpha} \left(\left(e^2 \epsilon^2 - \frac{\kappa^2}{4 e_h^2}\right) A_\mu \partial_\nu A_\alpha + e_h^2 \tilde{C}_\mu \partial_\nu \tilde{C}_\alpha\right) \ .
\label{get}
\end{split}
\end{equation}
Given Lagrangian \eqref{get}, the field equations read
\be
D_\mu[A]D^\mu[A] \phi + {\delta V}\!/{\delta \phi^*} = 0 \ , \label{nose}
\ee
\be
\frac{1}{12\pi m_1}\partial_\mu F^{\mu\nu}[A] +\frac{1}{4\pi} \left(1 - \frac{\kappa^2}{4 \epsilon^2 e^2 e_h^2}\right) \varepsilon^{\alpha\beta\nu} F_{\alpha\beta}[A] = -j^\nu \ , \label{eq:22}
\ee
\be
\varepsilon^{\alpha\beta\nu} {F}_{\alpha\beta}[\tilde C] = 0 \ ,
\label{eq:33}
\ee
where the current $j^\nu$ is defined as
\be
j^\nu = \vert\phi\vert^2 A^\nu + \frac{i}{2\epsilon e} \left(\phi \partial^\nu \phi^* - \phi^* \partial^\nu \phi\right) \ .
\ee

We see that the first two equations corresponds to a Maxwell-Chern-Simons-Higgs system completely decoupled from the $\tilde C_\mu$ field. Concerning $\tilde C_\mu$, Eq.~\eqref{eq:33} implies that it can be gauged away and then one has, using Eq.~\eqref{resta},
\be
C_\mu =- \frac{\kappa}{2 e_h^2} A_\mu \ .
\label{suma}
\ee
The associated magnetic $B_C$ and electric $E_C$ fields can then be related to those of the $ A_\mu$ field,
\be
B_A = -\frac{2 e_h^2}{\kappa}B_C \quad , \quad \vec{E}_A = -\frac{2 e_h^2}{\kappa} \vec{E}_C \ .
\label{trece}
\ee
and hence the associated magnetic fluxes $\Phi_A $ and $\Phi_C $ should take the form
\be
\Phi_A = \int B_A d^2x = \frac{2\pi n}{\epsilon e} = - \frac{2 e_h^2}{\kappa} \int B_C d^2x = - \frac{2 e_h^2}{\kappa} \Phi_C
\ee
Finally, we can deduce from \eqref{eq:22} that
\be
\label{eq:relacion}
\frac{1}{12\pi m_1} \vec \nabla\cdot \vec{E}_A + \frac{1}{2\pi} \left(1 - \frac{\kappa^2}{4 \epsilon^2 e^2 e_h^2}\right) B_A = j^0 = \lvert\phi\rvert^2 A_0 \ .
\ee
As it is always the case when a Chern-Simons term is present, there is a relation between magnetic fluxes $\Phi $ and electric charges $Q$, in this case given by
\be
Q_A =\frac{1}{2\pi} \left(1 - \frac{\kappa^2}{4 \epsilon^2 e^2 e_h^2}\right) \Phi_A \quad , \quad Q_C = -\frac{\kappa}{2e_h^2} Q_A\ .
\label{eq:charges}
\ee
Note that when $\kappa$ takes a critical value $\kappa_c$,
\be
\kappa_c = 2\epsilon e e_h \label{Critical}
\ee
both electric charges vanish. This is due to the fact that, at $\kappa = \kappa_c$, the Chern-Simons contribution for the $A_\mu$ gauge field equation \eqref{eq:22} vanishes and, as it is well known, in that case there are not finite energy electrically charged solutions.


In order to find vortex like solutions we propose the usual axially symmetric ansatz
\begin{align}
&A_\varphi = -\frac{a(r)}{r} \ , \ A_0 = \, a_0(r) \ , \\
&C_\varphi = -\frac{c(r)}{r} \ , \ C_0 = \, c_0(r) \ , \\
&\phi = \phi_0 f(r) e^{in\varphi}
\label{align}
\end{align}
which yields to the field equations
\begin{align}
&f'' + \frac{f'}{r} - \frac{f}{r^2}(n-\epsilon e a)^2 + \epsilon^2 e^2 f a_0^2 -\frac{\lambda \phi_0^2}{2} f (f^2 - 1) = 0 \ , \label{r} \\
&a'' -\frac{a'}{r} + \frac{12\pi m_1}{\epsilon e} \phi_0^2 f^2 (n-\epsilon e a) - 6m_1 \left(1 - \frac{\kappa^2}{4 e_h^2 \epsilon^2 e^2}\right) a_0' r = 0 \ , \label{s}\\
&a_0'' + \frac{a_0'}{r} - 12\pi m_1 \phi_0^2 a_0 f^2 - 6m_1 \left(1 - \frac{\kappa^2}{4 e_h^2 \epsilon^2 e^2}\right) \frac{a'}{r} = 0 \ , \label{t}\\
&c' + \frac{\kappa}{2 e_h^2} a'= 0 \ , \label{u}\\
&c_0' + \frac{\kappa}{2 e_h^2} a_0' = 0 \ . \label{v}
\end{align}
As for the boundary conditions, we set
\be
f(0) = a(0) = c(0) = 0 \ ,
\ee
\be
f(\infty) = 1 \quad , \quad a(\infty) = \frac{n}{\epsilon e} \quad , \quad a_0(\infty) = 0 \ .
\ee
We have not set conditions for $a_0$ at the origin, since its behaviour is constrained by equations \eqref{r}-\eqref{t} (see \cite{Jacobs}).

The electric and magnetic fields for the $A_\mu$ gauge field take the form
\be
\vec{E}_A = - a_0'\check{r} \quad , \quad B_A = -\frac{a'}{r} \ .
\ee
(Here $\check{r}$ is the radial unit vector). From this we can write the energy in the form
\begin{align}
	\mathcal E &= 2\pi \int{r \text{d} r\left(\frac{\epsilon^2 e^2}{24\pi m_1}\left(B_A^2 + E_A^2\right) -\frac{1}{2} \vert D_\mu[A]\phi \vert^2 + V[\phi]\right)}\\
	&\begin{multlined}
	= 2\pi \int{r \text{d} r\left(\frac{\epsilon^2 e^2}{24\pi m_1}\left(B_A^2 + E_A^2\right) + \frac{\phi_0^2}{2} \left(f'^2 + \frac{f^2}{r^2} (n-\epsilon e a)^2\right) + \right.}\\
	\left.\frac{\epsilon^2 e^2\phi_0^2}{2}a_0^2 f^2 + \frac{\lambda \phi_0^4}{4} \left(f^2-1\right)^2\right) \nonumber
	\end{multlined} \\
	&\equiv \int{\varepsilon(r) r \mathrm{d}r} \ ,
\end{align}
where $\varepsilon(r)$ is the energy density. It is interesting to note that there is no hidden sector contribution to the energy. This is due to the fact that the $CS[C]$ term is metric independent and hence its $T_{00}=0$ and since there is no hidden scalar there is no $C_0$ contribution through the Gauss law.

\section*{Asymptotic behaviour}

For large $r$, it can be shown that the fields behave as
\bea
a(r) &\to& \frac{n}{\epsilon e} + \alpha_\pm \phi_0 \sqrt{r} e^{-\mu_\pm r} \ ,\\
a_0(r) &\to& \pm \frac{\alpha_\pm}{\phi_0} \frac{e^{-\mu_\pm r}}{\sqrt{r}} \ ,\\
f(r) &\to& 1 + \frac{\beta_\pm}{\phi_0} \frac{e^{-m_h r}}{\sqrt{r}} \ ,\label{eq:36}
\eea
where $\alpha_\pm$ and $\beta_\pm$ are dimensionless constants, $m_h \equiv 1/\zeta = \phi_0\sqrt{\lambda/2}$ is the Higgs mass (with $\zeta$ the coherence length), and
\be
\mu_\pm = \pm\frac{\mu}{2} + \sqrt{\frac{\mu^2}{4} + 12 \pi m_1 \phi_0^2}
\label{38}
\ee
are the gauge field masses, with
\be
\label{mu}
\mu \equiv 3 m_1 \left(1 - \frac{\kappa^2}{4e_h^2\epsilon^2 e^2}\right) \ .
\ee
As it is well known, when the Higgs field and gauge field masses coincide, the abelian Higgs model can be supersymmetrized \cite{deVS}, both when the gauge field dynamics is governed by a Maxwell or a Chern-Simons term. This is not the case when both terms are present, as it is the case of Lagrangian \eqref{10vis}. However, let us note that in the low energy limit one can disregard the Maxwell term, and in that case supersymmetry would be achieved whenever
\be
\eta_\pm = \frac{m_h}{\mu_\pm} = 1 \ .
\ee
Let us also note that $\eta_\pm \geq 1$ corresponds to type II superconductivity, which is equivalent in the present case to
\be
1 \geq \frac{\kappa^2}{4e_h^2 \epsilon^2 e^2} \geq 1 - \sqrt{\frac{2}{\lambda}}\phi_0\left(\frac{\lambda}{6m_1}-4\pi\right)
\ee
for the $\mu_+$ case, whereas for $\mu_-$ we have
\be
1 \leq \frac{\kappa^2}{4e_h^2 \epsilon^2 e^2} \leq 1 + \sqrt{\frac{2}{\lambda}}\phi_0\left(\frac{\lambda}{6m_1}-4\pi\right) \ .
\ee
For each inequality, these expressions define a hypersurface of phase transition in the space of the present parameters, as can be found in \cite{Jacobs} for the Maxwell Chern-Simons Higgs model.


\section*{Numerical results}

We have studied numerically Eqs.~\eqref{r}-\eqref{v} in the type II superconductivity region using a second-order central finite difference procedure with an accuracy of ${\cal O}(10^{-4})$. We tested the solver for the bosonic sector of the action \eqref{10bis} setting $\kappa=0$, and accurately reproduced the exact result of the Bogomol'nyi lowest bound for the lowest-energy $n=1$ vortex, $\mathcal E = \phi_0^2\pi$ \cite{BPS,deVS}.

\begin{figure}[!htb]
\centering
\begin{subfigure}[b]{0.49\textwidth}
\includegraphics[width=\textwidth]{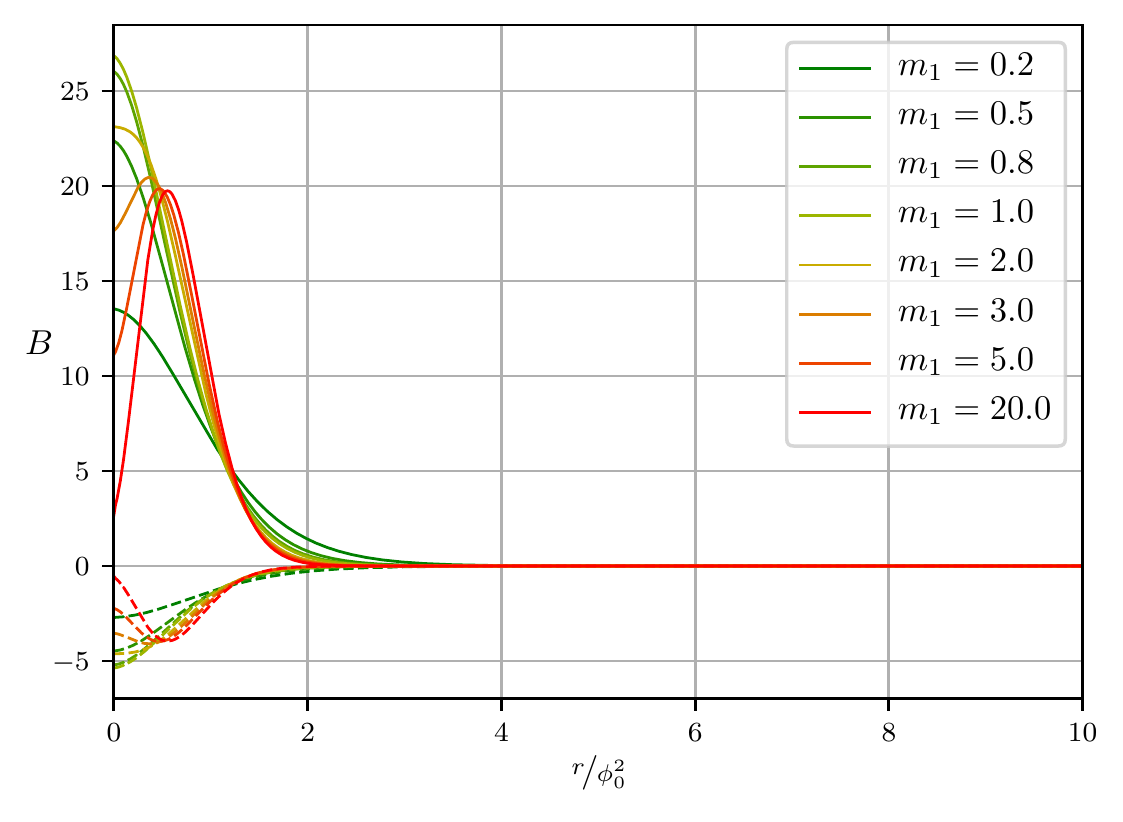}
\caption{}
\label{fig:Bm}
\end{subfigure}
\begin{subfigure}[b]{0.49\textwidth}
\includegraphics[width=\textwidth]{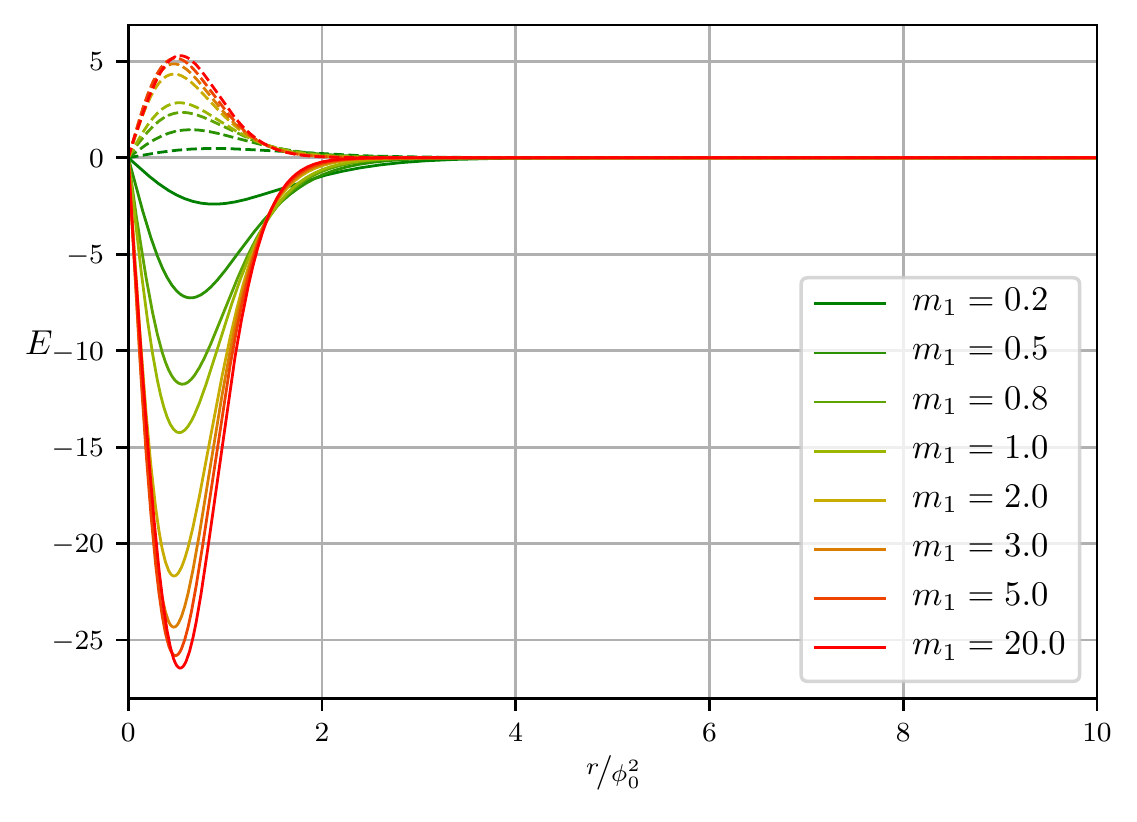}
\caption{}
\label{fig:Em}
\end{subfigure}
\caption{Profiles of magnetic and radial electric fields for the $A_\mu$ and $C_\mu$ gauge fields (dashed and solid lines, respectively) as $m_1$ changes, in units of $\phi_0^2$. The parameters were set to: $n = 1$, $\phi_0 = 1.0$, $\kappa = 0.1$, $e = 1.0$, $e_h=0.1$, $\lambda = 0.25$, $\epsilon=0.4$.}
\end{figure}

We studied the behaviour of the resulting profiles for the magnetic, electric and Higgs fields, and the energy density, for different values of the parameters of the model. In Figs.~\ref{fig:Bm}-\ref{fig:Em} we show the dependence of vortex-like magnetic fields ($B_A$ and $B_C$) and electric fields ($E_A$ and $E_C$) on $m_1$, the fermion $\psi^1$ mass, leaving the other parameters fixed. Note that, for fixed $\kappa$ and $e_h$, $B_A$ and $B_C$ are proportional and with opposite signs, according to Eq.~\eqref{trece}, as it is seen in the figures, and the same happens for $E_A$ and $E_C$. We observe that the maximum of $B_C$ increases as $m_1$ does for small values of $m_1$, whereas it decreases as $m_1$ increases for larger values. For large masses $B_C$ tends to ever smaller values at the origin, which is consistent with the fact that the magnetic field vanishes at $r=0$ for a pure Chern-Simons model. Indeed, making $m_1\to\infty$ implies dropping the Maxwell term in our original Lagrangian. 

\begin{figure}[!htb]
  \centering
  \begin{subfigure}[b]{0.49\textwidth}
  \includegraphics[width=\textwidth]{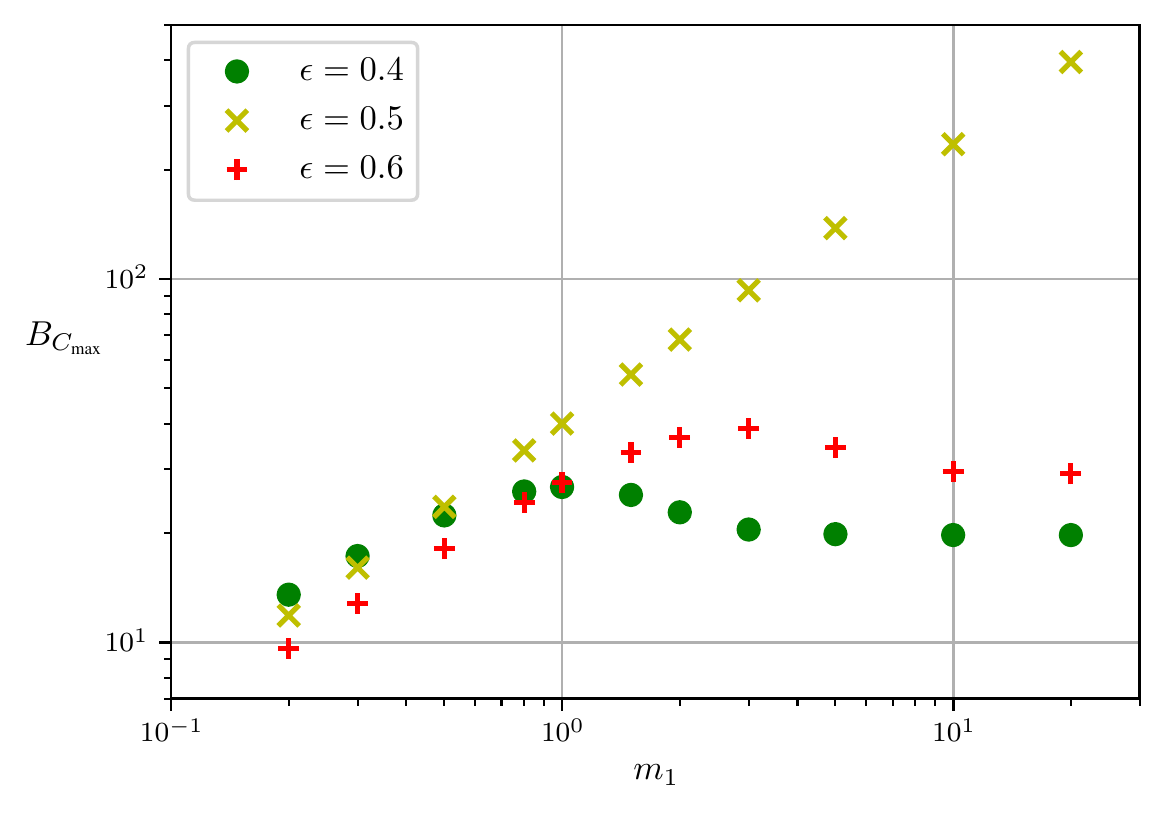}
  \caption{}
  \label{fig:maximo_B}
  \end{subfigure}
  \begin{subfigure}[b]{0.49\textwidth}
  \includegraphics[width=\textwidth]{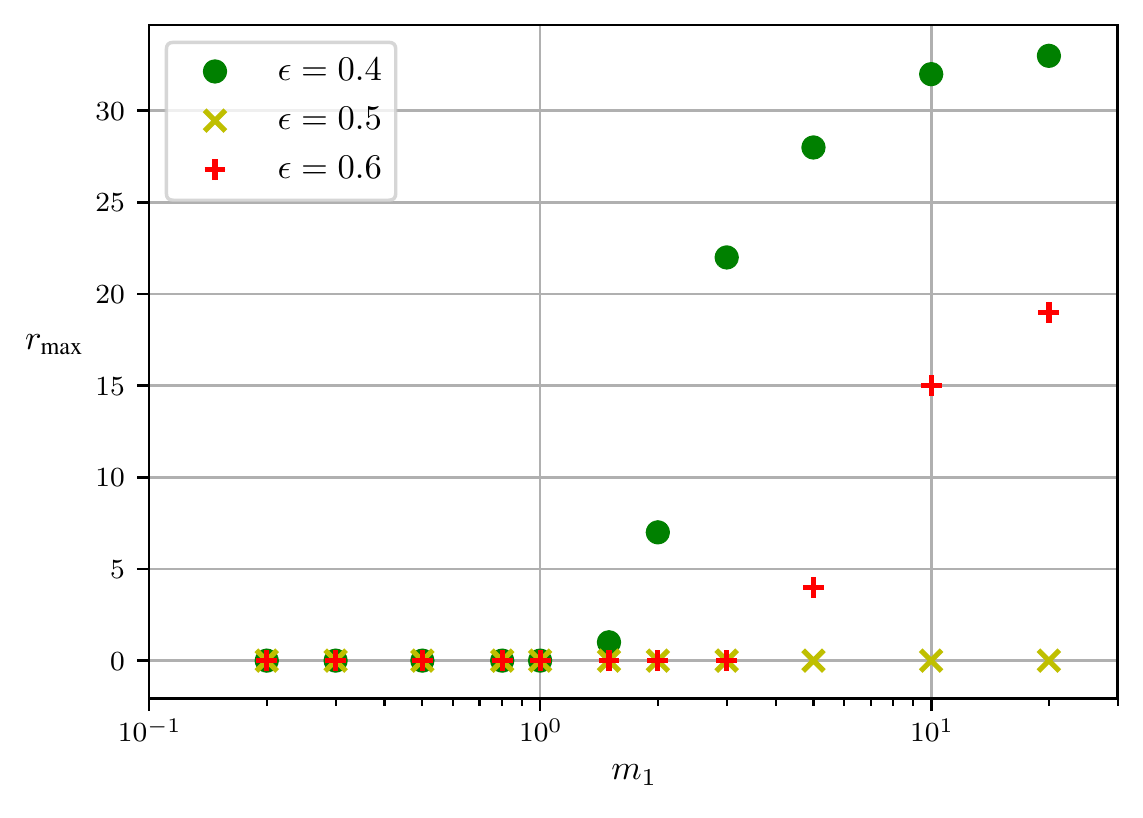}
  \caption{}
  \label{fig:maximo_B_r}
  \end{subfigure}
  \caption{Maximum values of $B_C$ (a) and the value $r_\text{max}$ where it is attained (b) as a function of $m_1$, for three different values of $\epsilon$. The parameters were set to: $n = 1$, $\phi_0 = 1.0$, $\kappa = 0.1$, $e = 1.0$, $e_h=0.1$, $\lambda = 0.25$.}
\end{figure}

Concerning  small masses, the Maxwell term becomes dominant and vortex solutions are similar to those for the Abelian Higgs model, with the  magnetic field maximum  tending to $r=0$. For a given value of $m_1$ (which could be estimated numerically) there is a transition and the maximum departs from the origin. It should be noted that, for the chosen values of the rest of the parameters of the model, this behavior holds for all different values of $\epsilon$ in the studied range $[0.1,1]$, except for the case $\epsilon = 0.5$. According to Eq.~\eqref{mu}, $\epsilon=0.5$ implies $\mu=0$, in which case the effective Lagrangian~\eqref{get} becomes the Abelian Higgs model Lagrangian, whose magnetic field attains its maximum at $r=0$ as stated above, no matter how large $m_1$ might be. Indeed, this behaviour can be seen in Figs.~\ref{fig:maximo_B}-\ref{fig:maximo_B_r}, showing the maximum of $B_C$ and the value $r_\text{max}$ where it is attained, as a function of $m_1$ and for three different values of $\epsilon$. It can be seen that, for $\epsilon \neq 0.5$, this maximum decreases for both large and small values of $m_1$, whereas it keeps increasing along with $m_1$ for the case $\epsilon=0.5$. Moreover, for $\epsilon\neq0.5$, we see that $r_\text{max}$ departs from from the origin $r=0$ for some value of $m_1$ (depending on $\epsilon$).

The qualitative behaviour of $E_C$, on the other hand, does not differ much from the usual pure CS model. For $m_1\to0$, $E_C$ vanishes since the Maxwell contribution dominates the Chern-Simons one and, as it is well known, there is no finite energy solution with non-zero electric field for the abelian Higgs model without a Chern-Simons contribution. The solution is the well-honoured Nielsen-Olesen vortex for the magnetic field.

\begin{figure}[!htb]
\centering
\begin{subfigure}[b]{0.49\textwidth}
\includegraphics[width=\textwidth]{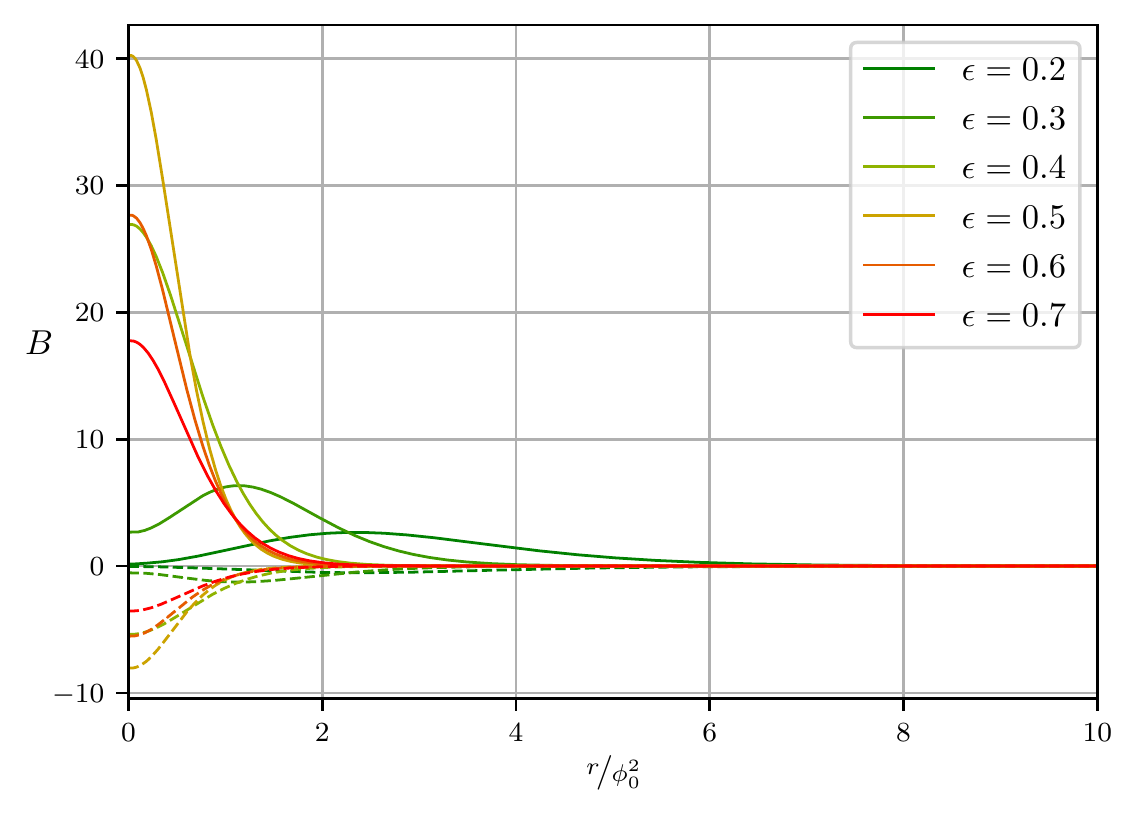}
\caption{}
\label{fig:Beps}
\end{subfigure}
\begin{subfigure}[b]{0.49\textwidth}
\includegraphics[width=\textwidth]{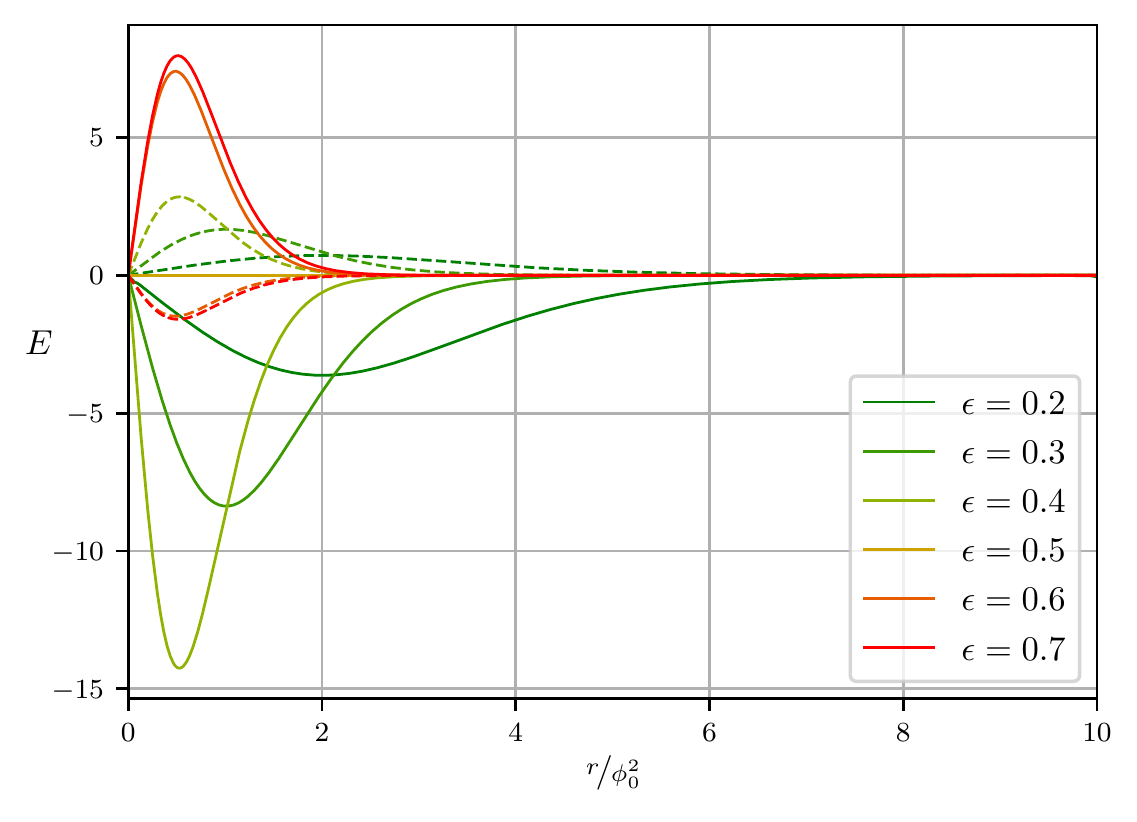}
\caption{}
\label{fig:Eeps}
\end{subfigure}
\caption{Profiles of magnetic and radial electric fields for the $A_\mu$ and $C_\mu$ gauge fields (dashed and solid lines, respectively) as $\epsilon$ changes. The parameters were set to: $n = 1$, $\phi_0 = 1.0$, $\kappa = 0.1$, $e = 1.0$, $e_h=0.1$, $\lambda = 0.25$, $m_1=1.0$.}
\end{figure}

In Figs.~\ref{fig:Beps}-\ref{fig:Eeps} we present the field profiles for different values of the visible coupling $\epsilon$. Contrary to what happens with $m_1$, the maximum of $B_C$ increases as $\epsilon$ does for large values of $\epsilon$, whereas it decreases as $\epsilon$ increases for smaller values. However, for small values of $\epsilon$ the field spreads and becomes less localized, whereas for large values of $m_1$ it remains localized around the same region. This behaviour was to be expected since, for $\epsilon\to0$, $A_\mu$ decouples from the $\psi^1$ fermions and the Higgs field $\phi$, and there is no $A_\mu$ dynamics. According to the effective Lagrangian \eqref{get}, all that remains is a Chern-Simons term for $A_\mu$, and hence $F_{\mu\nu}[A]=F_{\mu\nu}[C]=0$. This phenomenon also arises for the electric field, as it can be seen in Fig.~\ref{fig:Eeps}.

As for the electric field $E_C$, it can be seen in Fig.~\ref{fig:Eeps} that there is a change in its sign when passing from $\epsilon < 0.5$ to $\epsilon > 0.5$. According to Eq.~\eqref{mu} and the values of the other parameters, this implies a change in sign for $\mu$. As stated above, for $\mu=0$ the effective Lagrangian~\eqref{get} becomes the abelian Higgs model Lagrangian, which has zero electric field solution, as the figure shows.

Concerning the Higgs field, its behaviour does not depend considerably on the model parameters. Moreover, its profiles coincide qualitatively with those of the Nielsen-Olesen vortices, vanishing at the vortex core and reaching its VEV for larger $r$.

\begin{figure}[!htb]
\centering
\includegraphics[width=0.5\textwidth]{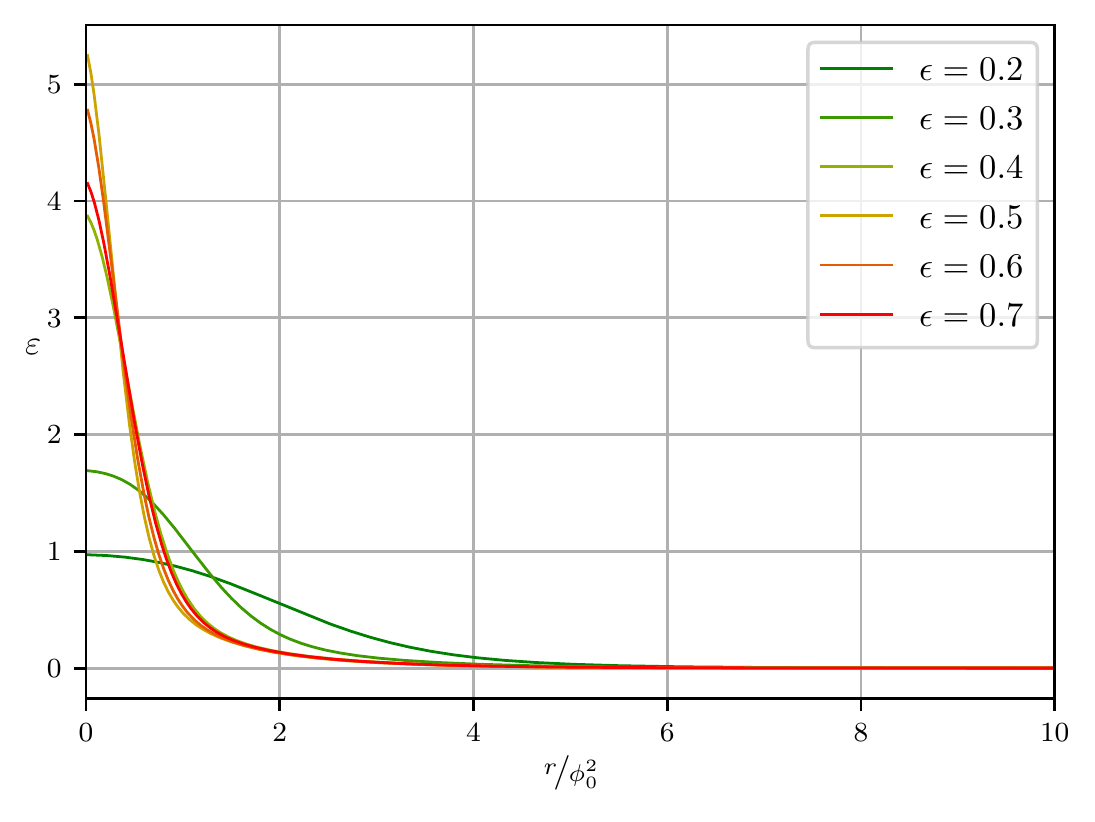}
\caption{Energy density $\varepsilon(r)$ as $\epsilon$ changes. The parameters were set to: $n = 1$, $\phi_0 = 1.0$, $\kappa = 0.1$, $e = 1.0$, $e_h=0.1$, $\lambda = 0.25$, $m_1=1.0$.}
\label{fig:energia_eps}
\end{figure}

As for the energy density, it is interesting to analyze its behaviour as $\epsilon$ changes. One can see in Fig.~\ref{fig:energia_eps} that, for small values of $\epsilon$, the concentration of energy spreads further away from the origin, consistently with the analogous behaviour of the electric and magnetic fields.

The results described above were obtained using the overal $+$ sign in the ambiguity of the Chern-Simons term arising from the regularization of determinants in odd dimensions. Choosing the overall negative sign yields a change in the sign of the constant $\mu$ (see Eqs.~\eqref{38}-\eqref{mu}) and in turn a change of the sign of the electric fields $E_A$ and $E_C$. It should be also mentioned that  the curves shape are similar to those corresponding to the $+$ sign discussed above. It should be also mentioned that this analysis for different values of $\epsilon$ was also carried out for other values of $m_1$, yielding qualitatively similar results, without any particular case for any value of $m_1$ (as opposed to the previous analysis for different values of $m_1$, where the case for $\epsilon=0.5$ was particularly different).

\FloatBarrier

\section*{Summary and discussion}
We have studied in this work a $2+1$ dimensional model in which the gauge and matter fields in a hidden sector are coupled to an external gauge field in the visible sector. Integration over fermion fields yielded an effective Lagrangian $L_{\text{eff}} $ \eqref{10vis} where instead of fermions one has Chern-Simons terms for both gauge fields and also Maxwell terms corrections at order $1/m$.

Proposing a Nielsen-Olesen like ansatz we studied the field equations associated to $L_{\text{eff}}$ and found vortex-like solutions for both gauge fields when the masses $m_1$ and $m_2$ of each of the two fermions are such that $m_2 \gg m_1$.

Interestingly enough, despite that the $U(1)$ symmetry associated to the $A_\mu$ gauge field has no explicit symmetry breaking potential, vortex-like solutions for this field with quantized $B_A$ flux do exist because of the mixing among both sectors. Analogously, there was no coupling between $C_\mu$ and and the symmetry breaking potential $V[\phi]$ and yet a non-trivial flux was found for the magnetic field $B_C$. Indeed,  ansatz \eqref{align} for the scalar field forces $A_\mu$ to have an asymptotic behavior $A_\varphi  \rightarrow n/(\epsilon e)$ since otherwise   the energy term associated to the covariant derivative $D_\varphi[A]$ would diverge. This the reason why there is a quantized $B_A$ flux, Then, since fields $A_\mu$ and $C_\mu$ are related according to Eq.~\eqref{suma}   also the $B_C$ flux is nontrivial quantized,

\be
\Phi_A = \int B_A d^2x = \frac{2\pi n}{\epsilon e} = - \frac{2 e_h^2}{\kappa} \Phi_C \ ,
\ee
and hence we see that $\Phi_C$ is also quantized, given by
\be
\Phi_C = \frac{\pi\kappa}{\epsilon e e_h^2}n \ .
\label{provided}
\ee
Let us note at this point that related models in which Chern-Simons terms  were included in one or both sectors have  been studied in refs.~\cite{f1}-\cite{f2}. However,  in those cases the idea was to consider  supersymmetric extensions that guarantee the existence of first order BPS equations, and of course this requires that gauge and scalar  fields should be present in both sectors. Moreover, as stated in the introduction, in the present case one can think of the $A_\mu$ gauge field  in the visible sector as an external field much like the role played by the external magnetic field in the phenomenological Landau-Ginzburg superconductivity theory, except that the order parameter is not in the visible sector.

As for the gauge symmetry breaking in a sector where there was no appropriate potential, it has been previously  discussed in ref.~\cite{f3} for the case of Maxwell gauge field dynamics   in both sectors and with the gauge mixing was of the $F_{\mu\nu}[A]F^{\mu\nu}[C]$ type. Even in the absence of a gauge symmetry breaking associated to the $A_\mu$ gauge field, a quantized $A_\mu$ magnetic flux was generated because of the gauge mixing. But since no Chern-Simons term  was present,  no electric charge could be generated neither in the visible nor in the hidden sector. In contrast, there exist electrically charged solutions, a result that could be relevant in the analysis of topological superconductors.

Another interesting result is that there is a critical value of the gauge mixing coupling constant $ \kappa = \kappa_c$ at which the vortex electric charges vanish since the field equations become those of a Maxwell-Higgs model which has no electrically charged finite energy solutions (see for example \cite{Manias}).

As mentioned in the introduction, there is a connection of the model we studied and those related to materials in which dualities in $2+1$ dimensions and topological order plays a central role. In particular, the effective dual theory discussed in reference \cite{MSen} can be related to ours if one takes the external gauge field $A_\mu$ as a probe with the dynamical fields describing the surface of topological insulators. We expect to discuss this issue thoroughly in a future work.


\vspace{1.2 cm}

\noindent{\bf{Acknowledgments:}}
F.A.S. is financially supported by PIP-CONICET,
PICT-ANPCyT and UNLP  grants.


\begin{thebibliography}{99}
\bibitem{Holdom}
  B.~Holdom,
  ``Two U(1)'s and Epsilon Charge Shifts'',
  Phys.\ Lett.\  {\bf 166B} (1986) 196.
\bibitem{uno}
 M.~Battaglieri {\it et al.},
 ``US Cosmic Visions: New Ideas in Dark Matter 2017: Community Report'',
 arXiv:1707.04591 [hep-ph].
 \bibitem{JR}
 J.~Jaeckel and A.~Ringwald,
 ``The Low-Energy Frontier of Particle Physics'',
 Ann.\ Rev.\ Nucl.\ Part.\ Sci.\ {\bf 60} (2010) 405.
 \bibitem{AriasNf} P.~Arias, E.~Ireson, C.~N\'u\~nez and F.~Schaposnik,
 ``$ \mathcal{N} $ =2 SUSY Abelian Higgs model with hidden sector and BPS equations'',
 JHEP {\bf 1502} (2015) 156.
 \bibitem{Ed} Steven Kivelson, Dung-Hai~Lee, Eduardo Fradkin and Vadim Oganesyan, ``Competing order in the mixed state of high temperature superconductors'' Phys.\ Rev.\ B {\bf 66} (2002) 144516.
 \bibitem{Frad} Eduardo Fradkin, Steven A. Kivelson, and John M. Tranquada.
``Theory of Intertwined Orders in High Temperature Superconductors'', Rev.\ Mod.\ Phys {\bf 87} (2015) 457.
\bibitem{Ba} Egor Babaev and Martin Speight, ``Semi-Meissner state and neither type-I nor type-II superconductivity in multicomponent
superconductors'', Phys.\ Rev.\ B {\bf 72} (2005) 180502 (R).
\bibitem{Zhang} S.-C.\ Zhang, ``A Unified Theory Based on $SO(5)$ Symmetry of Superconductivity and Antiferromagnetism'', Science {\bf 275} (1997) 1089.
\bibitem{Co} D. P.\ Arovas, A. J.\ Berlinsky, C.\ Kallin, and S.-C. Zhang, ``Superconducting Vortex with Antiferromagnetic Core'',  Phys.
Rev. Lett. {\bf 79} (1997) 2871.
 \bibitem{Sen} A.\ Vishwanath and T.\ Senthil, ``Physics of Three-Dimensional Bosonic Topological Insulators: Surface-Deconfined Criticality and Quantized Magnetoelectric Effect'',	Phys. Rev. X {\bf 3} (2013) 011016.
\bibitem{SSWW}
 N.~Seiberg, T.~Senthil, C.~Wang and E.~Witten,
 ``A Duality Web in 2+1 Dimensions and Condensed Matter Physics'',
 Annals Phys.\ {\bf 374} (2016) 395.
\bibitem{f3} P.~Arias and F.~A.~Schaposnik,
``Vortex solutions of an Abelian Higgs model with visible and hidden sectors'',
  JHEP {\bf 1412} (2014) 011.
 \bibitem{Wein0} S.~Weinberg,
 ``Goldstone Bosons as Fractional Cosmic Neutrinos'',
 Phys.\ Rev.\ Lett.\ {\bf 110} (2013) no.24, 241301.
 \bibitem{Arias}
 P.~Arias, E.~Ireson, C.~N\'u\~nez and F.~Schaposnik,
 ``$ \mathcal{N} $ =2 SUSY Abelian Higgs model with hidden sector and BPS equations'',
 JHEP {\bf 1502} (2015) 156.
 \bibitem{Gru} D.\ Savchenko and A.\ Rudakovskyi, ``New mass bound on fermionic dark matter from a combined analysis of classical dSphs'', Mon.\ Not.\ Roy.\ Astron. Soc. {\bf 487} (2019) 5711.

 \bibitem{Argo}
 R.~Acciarri {\it et al.} [ArgoNeuT Collaboration],
 ``Improved Limits on Millicharged Particles Using the ArgoNeuT Experiment at Fermilab'',
 arXiv:1911.07996 [hep-ex].


\bibitem{Wj} L.~C.~R.~Wijewardhana,
 ``Induced Chern-simons Terms At High Temperature'',
 Physica A {\bf 158} (1989) 33.
 \bibitem{JaLibro} R.\ Jackiw, Diverse Topics In Theoretical And Mathematical Physics, World Sci., Singapore, 1995.
     \bibitem{Hamed}
  N.~Arkani-Hamed, D.~P.~Finkbeiner, T.~R.~Slatyer and N.~Weiner,
  ``A Theory of Dark Matter'',
  Phys.\ Rev.\ D {\bf 79} (2009) 015014.
\bibitem{BPS}E.~B.~Bogomolny,
 ``Stability of Classical Solutions'',
 Sov.\ J.\ Nucl.\ Phys.\ {\bf 24} (1976) 449
 [Yad.\ Fiz.\ {\bf 24} (1976) 861].
\bibitem{deVS} H.~J.~de Vega and F.~A.~Schaposnik,
 ``A Classical Vortex Solution of the Abelian Higgs Model'',
 Phys.\ Rev.\ D {\bf 14} (1976) 1100.
\bibitem{R} M.~Ahlers, H.~Gies, J.~Jaeckel, J.~Redondo and A.~Ringwald,
 ``Laser experiments explore the hidden sector'',
 Phys.\ Rev.\ D {\bf 77} (2008) 095001.

 \bibitem{Redlich} A.~N.~Redlich,
 ``Gauge Noninvariance and Parity Violation of Three-Dimensional Fermions'',
 Phys.\ Rev.\ Lett.\ {\bf 52} (1984) 18; ``Parity Violation and Gauge Noninvariance of the Effective Gauge Field Action in Three-Dimensions'',
 Phys.\ Rev.\ D {\bf 29} (1984) 2366.
 \bibitem{GRS} R.~E.~Gamboa Saravi, G.~L.~Rossini and F.~A.~Schaposnik,
 ``The Zeta function answer to parity violation in three-dimensional gauge theories'',
 Int.\ J.\ Mod.\ Phys.\ A {\bf 11} (1996) 2643.
 \bibitem{Wein} S. Weinberg, The Quantum Theory of Fields. Vol. 1:
Foundations (Cambridge University Press, Cambridge,
England, 1995).
\bibitem{MS}
 E.~F.~Moreno and F.~A.~Schaposnik,
 ``Dualities and bosonization of massless fermions in three dimensional space-time'',
 Phys.\ Rev.\ D {\bf 88} (2013) 025033.
 \bibitem{PS} P.~Arias and F.~A.~Schaposnik,
 ``Vortex solutions of an Abelian Higgs model with visible and hidden sectors'',
 JHEP {\bf 1412} (2014) 011.
 \bibitem{Jacobs}
 L.~Jacobs, A.~Khare, C.~Nagaraja Kumar and S.~K.~Paul,
 ``The Interaction of {Chern-Simons} Vortices'',
 Int.\ J.\ Mod.\ Phys.\ A {\bf 6} (1991) 3441.
 \bibitem{PIST}
 J.~M.~P\'erez Ipi\~na, F.~A.~Schaposnik and G.~Tallarita,
 ``SU(2) Chern-Simons Theory Coupled to Competing Scalars'',
 Phys.\ Rev.\ D {\bf 97} (2018) no.11, 116010.
 \bibitem{f1}P.~Arias, E.~Ireson, F.~A.~Schaposnik and G.~Tallarita,
  ``Chern-Simons-Higgs theory with visible and hidden sectors and its N=2 SUSY extension'',
  Phys.\ Lett.\ B {\bf 749} (2015) 368.
 \bibitem{f2} E.~Ireson, F.~A.~Schaposnik and G.~Tallarita,
  ``Visible and hidden sectors in a model with Maxwell and Chern-Simons gauge dynamics'',
  Int.\ J.\ Mod.\ Phys.\ A {\bf 31} (2016) no. 34, 1650178.
\bibitem{Manias}
 M.~V.~Manias, C.~M.~Naon, F.~A.~Schaposnik and M.~Trobo,
 ``Nonabelian Charged Vortices As Cosmic Strings'',
 Phys.\ Lett.\ B {\bf 171} (1986) 199.
 \bibitem{MSen} M. A.\ Metlitski and A.\ Vishwanath, ``Particle-vortex duality of 2d Dirac fermion from electric-magnetic duality of 3d topological insulators'', Phys.\ Rev.\ B {\bf 93}, (2016) 245151.
\end{thebibliography}
\end{document}